%% file: main.tex
\documentclass[11pt]{article}

\usepackage{amsthm}
\usepackage{amsmath}
\usepackage{amssymb}
\usepackage{amsmath, amsthm, bbm}
\usepackage{framed, fullpage}
\usepackage{algorithm}
\usepackage[backref,colorlinks,citecolor=blue,bookmarks=true]{hyperref}
\usepackage{color,multicol}
\usepackage[dvipsnames]{xcolor}

\def\conf{0}

\newcommand{\s}[1]{\left\lvert #1 \right\rvert}

\newcommand{\e}{\epsilon}

\newcommand{\sz}{s}  
\renewcommand{\d}{{\rm dist}} 

\newcommand{\cent}{\sigma}
\newcommand{\W}{W}  
\newcommand{\cell}{C}  
\newcommand{\Path}{P}  
\newcommand{\chiW}{\chi_{\mbox{\tiny\it W}}}

\renewcommand{\Pr}{\mathrm{Pr}}
\newcommand{\Exp}{\mathrm{Exp}}
\newcommand{\Var}{\mathrm{Var}}
\newcommand{\cov}{\mathrm{Cov}}

\newtheorem{fact}{Fact}

\newcommand{\eps}{\epsilon}
\newcommand{\eqdef}{\stackrel{\rm def}{=}}
\newtheorem{defn}{Definition}         
\newcommand{\BD}{\begin{defn}} \newcommand{\ED}{\end{defn}}
\newcommand{\BE}{\begin{enumerate}} \newcommand{\EE}{\end{enumerate}}
\newcommand{\BI}{\begin{itemize}} \newcommand{\EI}{\end{itemize}}
\newcommand{\calA}{{\cal A}}

\newcommand{\calT}{{\cal T}}

\newcommand{\calP}{{\cal P}}
\newtheorem{thm}{Theorem}
\newcommand{\BT}{\begin{thm}} \newcommand{\ET}{\end{thm}}
\def\FullBox{\hbox{\vrule width 8pt height 8pt depth 0pt}}
\newcommand{\ourqed}{\;\;\;\FullBox}
\newenvironment{ourproof}{\noindent{\bf Proof:~~}}{\(\ourqed\)}
\newcommand{\BPF}{\begin{ourproof}} \newcommand {\EPF}{\end{ourproof}}
\newenvironment{proofof}[1]{\noindent{\bf Proof of {#1}:~~}}{\(\ourqed\)}
\newcommand{\BPFOF}{\smallskip \begin{proofof}} \newcommand {\EPFOF}{\end{proofof}}
\newcommand{\BEQ}{\begin{equation}} \newcommand{\EEQ}{\end{equation}}
\newcommand{\BEQN}{\begin{eqnarray}}\newcommand{\EEQN}{\end{eqnarray}}
\newtheorem{lem}{Lemma}      
\newcommand{\BL}{\begin{lem}} \newcommand{\EL}{\end{lem}}
\newtheorem{clm}[lem]{Claim}
\newcommand{\BCM}{\begin{clm}} \newcommand{\ECM}{\end{clm}}
\newtheorem{cor}[thm]{Corollary}      
\newcommand{\BC}{\begin{cor}} \newcommand{\EC}{\end{cor}}
\newcommand{\poly}{{\rm poly}}
\newcommand{\BF}{\begin{fact}} \newcommand{\EF}{\end{fact}}

\newcommand{\mnote}[1]{{\color{red}$\spadesuit$} \marginpar{\tiny\bf
            \begin{minipage}[t]{0.5in}
              \raggedright #1
         \end{minipage}}}

\newcommand{\dnew}[1]{{#1}}
\newcommand{\ddnew}[1]{{#1}}

\newcommand{\rrnew}[1]{{#1}}
\newcommand{\old}[1]{{#1}}

\date{}

\title{A Local Algorithm for Constructing Spanners in Minor-Free Graphs}
\author{
Reut Levi\thanks{MPI for informatics, Saarbr\"{u}cken 66123, Germany.
  Email: {\tt rlevi@mpi-inf.mpg.de}.}
\and
Dana Ron\thanks{School of Electrical Engineering, Tel Aviv University.
  Tel Aviv 69978, Israel.
  Email: {\tt danar@eng.tau.ac.il}.
This research was partially supported by the Israel Science Foundation grant No.~671/13}
\and
Ronitt Rubinfeld\thanks{CSAIL, MIT.
  Cambridge MA 02139, USA.
Blavatnik School of Computer Science, Tel Aviv University.
  Tel Aviv 69978, Israel.
  Email: {\tt ronitt@csail.mit.edu}.
This research was partially supported by the NSF grant CCF-1420692 and ISF grant 1536/14}
}

\begin{document}

\begin{titlepage}
\maketitle
\thispagestyle{empty}

\begin{abstract}

Constructing a spanning tree of a graph is
one of the most basic tasks in graph theory.
 We consider this problem in the setting of local algorithms:
one wants to quickly determine whether a given edge $e$ is in a specific spanning tree,
without computing the whole spanning tree, but rather by inspecting
the local neighborhood of $e$.
The challenge is to maintain consistency. That is, to answer queries
about different edges according to
the {\em same\/} spanning tree.
Since it is known that this problem cannot be solved  without essentially
viewing all the graph, we consider the relaxed version
of finding a spanning subgraph 
with
$(1+\eps)n$ edges instead of $n-1$ edges
(where $n$ is the number of vertices and $\eps$ is a given approximation/sparsity parameter).

It is known that this relaxed problem requires inspecting $\Omega(\sqrt{n})$ edges in general graphs
(for any constant $\eps$),
which motivates the study of natural restricted families of graphs.
One such family is the family of graphs with an excluded minor (which in particular
includes planar graphs).
For this family there is an algorithm that
	 achieves constant success probability, and inspects
$(d/\eps)^{\poly(h)\log(1/\eps)}$ edges
 (for each edge it is queried on),
where $d$ is the maximum degree in the graph and $h$ is the size of the excluded minor.
The distances between pairs of vertices in the spanning subgraph $G'$ are at most
a factor of $\poly(d, 1/\eps, h)$ larger than in $G$.

In this work, we show that for an input graph that is $H$-minor free for any $H$ of size $h$, this task can be performed
by inspecting only $\poly(d, 1/\eps, h)$  edges in $G$.
The distances between pairs of vertices in the spanning subgraph $G'$ are at most
a factor of 
$\tilde{O}(h\log(d)/\eps)$
larger than in $G$.
Furthermore, the error probability of the new algorithm is significantly improved  to $\Theta(1/n)$.
This algorithm can also be easily adapted to yield an efficient algorithm for the distributed
(message passing) setting.
\end{abstract}

\end{titlepage}

\input{intro}

\input{local-spanning-minor-free}

\section{Acknowledgement}
We would like to thank Oded Goldreich for his helpful suggestions.
\bibliographystyle{plain}
\bibliography{refs}

\appendix
\section{A non-polynomial relation between $k$ and $Y_k(v)$}\label{app:Y-k-big}
Recall that $Y_k(v) \eqdef \{u\in V: v \in B_k(u) \}$.
In the following lemma we show that $|Y_k(v)|$ can be super polynomial.
\BL\label{lem:Y-k-big}
There exists a graph $G=(V,E)$ with degree bounded by $d$ and $v\in V$ such that
$|Y_k(v)| = k^{\Omega(\log \log d)}$.
\EL
\BPF
The graph $G$ is a tree, rooted at $v$, and defined as follows.
For simplicity, we let the degree bound be $d+1$ (so that a vertex may have $d$ children, and
hence degree $d+1$).
We partition the levels of the tree into consecutive subsets: $L_0 = \{1,\dots,\ell_0\}$ (where the
root is at level $1$),
$L_1 = \{\ell_0+1,\dots,\ell_1\}$, $\dots$, $L_r = \{\ell_{r-1}+1,\dots,\ell_r\}$.
For each subset $L_i$, and for each level $j$ in the subset, all vertices in level $j$ have the same number of children, which is $d_i \eqdef d^{2^{-i}}$. We set $r = \log\log d$, so that all vertices in
levels belonging to $L_r$ have two children.
Finally we set $s_i = |L_i| = \log_{d_i} g^{1/(r+1)}$, where $g$  determines the size of the
tree, as well as the minimum $k$ that ensures that all vertices in the tree belong to $Y_k(v)$.

By the construction of the tree, the number of vertices in it is of the order of
$\prod_{i=0}^r d_i^{s_i} = g$. In order to upper-bound $k$ (such that all vertices belong to
$Y_k(v)$), consider any vertex $u$ in some level $j \in L_i$, where $0 \leq i \leq r$.
Since $d_i = d_{i-1}^{1/2}$, so that $s_t = 2 s_{t-1}$ for each $t$, we get that $s_i \leq \sum_{i' < i} s_{i'}$. Therefore,
$\d(u,v) \leq 2s_i$. It follows that the number of vertices in the subtree rooted at $u$
that are at distance at most $\d(u,v)$ from $u$ is upper bounded by
$d_i^{s_i}\cdot d_{i+1}^{s_i} < g^{3/2(r+1)}$. Since this is true for every vertex in the tree, we
get that $|\Gamma_{\d(u,v)}(u)| = O(g^{3/2(r+1)}\cdot s_r) = O(g^{3/2(r+1)}\cdot\log g)$,
which gives us an upper bound on $k$, from which the lemma follows.
\EPF

\ifnum\conf=1
\section{Analysis of the probe complexity and running time of Algorithm~\ref{alg:ssg}}\label{app:probe-ssg}
The number of vertices that Algorithm~\ref{alg:findc} inspects
(for any vertex $v$ it is called with) is at most $kd$. Since the degree of each vertex is bounded by $d$, its probe complexity is bounded by $kd^2$.
Algorithm~\ref{alg:findc} makes at most $kd$ accesses to $\chiW$, hence, by Theorem~\ref{thm:twise} its running time is bounded by $O(k^2d^2)$.
The probe complexity of Algorithm~\ref{alg:bfs} is upper bounded by the total probe complexity
of at most $d$ calls to  Algorithm~\ref{alg:findc}, plus $d$ executions of a BFS until at most
$kd$ vertices are encountered (Steps~\ref{st:u-up} and~\ref{st:up-w}, which for simplicity
we account for separately from Algorithm~\ref{alg:findc}).
A similar bound holds for the running time.
Hence, the total probe complexity and running time  of Algorithm~\ref{alg:bfs} are  $O(k d^3)$ and $O(k^2 d^3)$, respectively.

The size of any subtree returned by Algorithm~\ref{alg:treed} is upper bounded by $\sz^2  d^2$.
To verify this, recall that at the end of Step~\ref{step:bfs} of Algorithm~\ref{alg:treed}, at most $\sz d$ vertices were explored. Hence, the number of vertices that are incident to the explored vertices is at most  $\sz d^2$.
Thus, due to Step~\ref{step:check}, the total number of vertices in each part is at most $\sz^2 d^2$.
Since Step~\ref{step:check} can be implemented locally by calling Algorithm~\ref{alg:bfs} at most $\sz$ times, we obtain that the total probe complexity and running time of revealing each part is at most $O(\sz^2 k d^5)$ and $O(\sz^2 k^2 d^5)$, respectively.
The probe complexity and running time of Algorithm~\ref{alg:ssg} are dominated by those of Step~\ref{step:part}.
Observe that in Step~\ref{step:part} at most $|\Path(y)|$ parts are revealed.
Since $|\Path(y)|  \leq k$, we obtain that  the overall probe complexity and running time  of Algorithm~\ref{alg:ssg} are bounded by $O(\sz^2 k^2 d^5)$ and $O(\sz^2 k^3 d^5)$, respectively.
By the settings of $k$ and $\sz$ we obtain the final result.
\fi

\end{document}

%% file: intro.tex
\section{Introduction}


Given  graph  $G=(V,E)$, a basic task
is to find a sparse spanning subgraph $G'$, where $G'$ may be required to
be a tree, or may be required to approximately preserve distances between
vertices (i.e., have small {\em stretch\/}~\cite{PU89,PS89}).
Suppose we are interested in determining whether a given edge $e$
belongs to the spanning subgraph $G' = (V,E')$.  
We can of course run an algorithm that constructs $G'$ and check
whether $e \in E'$, but can we do better?
In particular, is it possible to answer such queries
with a number of operations that
is {\em sublinear} in $n=|V|$ or
possibly even independent of $n$?

This local version of the problem was first studied in~\cite{LRR14}.
Observe that the main challenge  is to answer
queries about different edges consistently with the same sparse spanning graph $G'$,
while being allowed to inspect only a small (local) part of the graph for each queried edge.
Moreover, while the underlying spanning graph may depend on the internal randomness of
the algorithm that answers the queries, it should not depend on the (order of the) queries themselves.
The following simple but important observation appears in~\cite{LRR14}.
If one insists that $G$ have a minimum number of edges, namely, that $G'$ be a tree,
then it is easy to see that
there are graphs $G$ that contain edges $e$ for which determining
whether $e \in G'$ requires inspecting a linear number of edges in $G$.
To see this, observe that
if $G$ consists of a single path, then the algorithm must answer positively on
all edges, while if $G$ consists of a cycle, then the
algorithm must answer negatively on one edge. However, the
two cases cannot be distinguished without inspecting a
linear number of edges.

Given the above observation, the question posed in~\cite{LRR14}
is whether a relaxed version of this task can be solved by
inspecting a sublinear number of edges, where the relaxation is
that the spanning graph $G'$ may contain $(1+\eps)\cdot n$ edges,
for a given approximation/sparsity parameter $\eps$.
As shown in~\cite{LRR14}, in general, even after this relaxation, local algorithms for sparse
subgraphs require the inspection of $\Omega(\sqrt{n})$ 
edges for each queried edge $e$ and for a constant $\eps$.
This lower bound holds for graphs with strong expansion properties, 
and there is an almost matching upper bound
for such graphs. In fact, even for graphs with relatively weak expansion properties, there is no
local algorithm that inspects a number of 
edges that is independent of $n$~\cite{LMRRS}. However, for graphs that are sufficiently non-expanding
there are local algorithms whose complexity is independent of
$n$ (depending only on the maximum degree $d$ and  the approximation parameter $\eps$)~\cite{LRR14,LMRRS}.

A family of non-expanding graphs that is of wide interest, is the family of {\em minor-free\/}
graphs, which can be parameterized by the size, $h$, of the excluded minor.
In particular, this family includes planar graphs.
It was shown in~\cite{LRR14} that, based on~\cite{LR15}, it is possible to obtain
a local sparse spanning tree algorithm for graphs that are free of a minor
of size $h$, where the complexity of the algorithm is $(d/\eps)^{\poly(h)\log(1/\eps)}$
  and which has  high constant success probability.

\input{results}

\input{related-work}

%% file: results.tex
\subsection{Our Result and Techniques}\label{subsec:result}
In this work we significantly improve on the aforementioned result
for the class of minor-free graphs,
by designing
a local sparse spanning graph algorithm for minor-free graphs whose complexity is polynomial in
$h$, $d$ and $1/\eps$ and which has success probability $1-1/\Omega(n)$.
We note that though graphs with excluded minors have bounded average degree,
the {\em maximum} degree of such graphs remains unbounded, and our local algorithms
have complexity that depends on this maximum degree.
The spanning graph  $G'$ maintains
the minimum distances between pairs of vertices in $G$ up to a factor
of $\tilde{O}(h\log(d)/\eps)$.
Namely, it has {\em stretch\/}~\cite{PU89,PS89}
$\tilde{O}(h\log (d)/\eps)$
(the stretch obtained in~\cite{LRR14} is somewhat higher, and in particular polynomial in $d$).

Similarly to some of the other local algorithms for sparse spanning graphs presented in~\cite{LRR14,LMRRS},
our algorithm is based on defining an underlying global partition (which is randomized). The global partition determines the sparse spanning graph, and the local algorithm decides whether  a given edge belongs to the spanning graph by constructing parts of the
partition. The differences between the algorithms are in
the way the partition is defined, the way the spanning graph is determined by the partition
(more specifically, the choice of edges between parts), and
the local partial construction of the partition.
Interestingly, our partition and its construction bear more similarities to
the underlying partition defined by the local sparse spanning
graph algorithm for highly expanding graphs, than
the partition used by the previous algorithm for minor-free graphs~\cite{LRR14}.
In what follows we provide a high level description of the partition, the sparse spanning graph it defines, and the corresponding local algorithm.
We also explain how we can directly obtain a distributed algorithm, and give
a bound on its round complexity.

\paragraph{A partition-based algorithm and the construction of $G'$.}
Our local algorithm is based on defining an underlying global partition of
the vertices into many small connected parts, where the
partition satisfies certain properties defined in the next paragraph.
Given such a partition, the edge set $E'$ of $G'$ is simply
defined by taking a spanning tree in
each part, and a single edge between every pair of parts that have
at least one edge between them.
The minor-freeness of the graph, together with a bound on the number of parts,
ensure that $|E'| \leq (1+\eps)n$.

\ifnum\conf=1
\vspace{-1ex}
\fi
\paragraph{Properties of the partition.}
We define a partition that has each of the four following
properties (which for simplicity are not precisely quantified in this
introductory text): (1) the number of parts in
the partition is not too large;
(2) the size of each part is not too large;
(3) each part induces a connected subgraph;
(4) for every vertex $v$ it is possibly to efficiently find the
subset of vertices in the part that $v$ belongs to.

\ifnum\conf=1
\vspace{-1ex}
\fi
\paragraph{An initial centers-based partition.}
Initially, the partition is defined by a selection of {\em centers\/}, where
the centers are selected randomly (though not completely independently).
Each vertex is initially assigned to the closest selected center. For an appropriate setting
of the probability that a vertex is selected to be a center, 
with high probability, this initial partition has the first property.
Namely, the number of parts in the partition is not too large.
By the definition of the partition, each part is connected, so that
the partition has the third property as well.
However,
the partition is not expected to have the second property,
that of each part having small size.  Even for a simple graph
such as the line, we expect to see parts of size logarithmic in $n$.
In addition, the same example shows that the
fourth property may not hold, i.e. there may be many vertices for which
it takes superconstant time to find the part that the vertex belongs to.
To deal with these two problems,
we give a procedure for refining the partition in two phases, as explained next.

\ifnum\conf=1
\vspace{-1ex}
\fi
\paragraph{Refining the partition, phase 1 (and a structural lemma).}
A first refinement  is obtained by the separation of vertices that in order to reach their assigned center
in a Breadth First Search (BFS), must observe a number of vertices that is above a
certain predetermined threshold, $k$.
Each of these {\em remote\/} vertices becomes a singleton
subset in the partition. We prove that with probability $1-1/\Omega(n)$, the number of
these remote vertices is not too large, so that the first property (concerning the number of parts)
is maintained with high probability.

The probabilistic analysis builds on
a structural lemma, which may be of independent interest.
The lemma upper bounds, for any given vertex $v$, the number of vertices $u$ such that
$v$ can be reached from $u$ by performing a BFS until at most $k$ vertices are observed.
This number in turn upper bounds the size of each part in the partition after the
aforementioned refinement.
While this upper bound is not polynomial in $k$ and $d$, it suffices for the
purposes of our probabilistic (variance) analysis (and we also show that there is
no polynomial upper bound).

\ifnum\conf=1
\vspace{-1ex}
\fi
\paragraph{Refining the partition, phase 2.}
In addition to the first property, the third property (connectivity of parts) is also
maintained in the resulting refined partition, and  the refinement
partially addresses the fourth property,
as remote vertices can quickly determine that they are far from all centers.
In addition, the new parts of the partition will be of size $1$ and thus will
not violate the second property.
However, after this refinement, there might be some large parts remaining
so that the second and fourth properties are  not necessarily satisfied.
We further partition large parts into smaller (connected) parts by a
certain decomposition based on the BFS-tree of the large part.

\ifnum\conf=1
\vspace{-1ex}
\fi
\paragraph{The local algorithm.}
Given an edge $\{u,v\}\in E$, the main task of the local algorithm is to find the two parts to which the vertices belong in the final partition. Once these parts are found, the algorithm can easily decide whether the edge between $u$ and $v$ should belong to $E'$ or not. In order to find the part
that $u$ (similarly, $v$) belongs to, the local algorithm does the following.
First it performs a BFS until it finds a center, or it sees at least $k$ vertices (none of which is a center).
In the latter case, $u$ is a singleton (remote) vertex. In the former case, the algorithm has found the
center that $u$ is assigned to in the initial partition, which we denote by $\cent(u)$. The algorithm next performs a BFS starting from $\cent(u)$. For each vertex that it encounters, it checks whether this vertex is assigned to
$u$ (in the initial partition). If the number of vertices that are found to be assigned to $\cent(u)$ is
above a certain threshold, then the decomposition of the part assigned to $\cent(u)$ needs to be emulated locally.
We show that this can be done recursively in an efficient manner.

\ifnum\conf=1
\vspace{-1ex}
\fi
\paragraph{A distributed  algorithm.}
Our algorithm easily lends itself to an implementation in the distributed (message passing)
setting. In this setting a processor resided on each vertex in the graph. Computation proceeds in
rounds, where in each round every vertex sends messages to its neighbors.
In the distributed implementation, initially each vertex decided, independently at random (and with
the appropriate probability), whether it is a center (of the initial partition).
In the first round, each vertex sends its id (name) to each of its neighbors, as well as
an indication whether it is a center. In the following rounds, each center send all its neighbors
the information it has gathered about its local neighborhood (including the identity of the centers).
It follows from our analysis that after $\tilde{O}(h\log(d)/\eps))$ rounds, each processor
can determine if its incident edges belong to the sparse spanning graph $G'$.

%% file: related-work.tex
\subsection{Related Work}\label{subsec:related}
As mentioned earlier, the works most closely related to the current work are~\cite{LRR14,LMRRS}.
In addition to the results in these works that were already described,
in~\cite{LRR14} there is a local sparse spanning graph algorithm for the family of $\rho$-hyperfinite graphs\footnote{A graph is
{\em $\rho$-hyperfinite\/} for a function $\rho: \mathbb{R}_+ \to \mathbb{R}_+$,
if its vertices can be partitioned into
subsets of size at most $\rho(\eps)$ that are connected and such that the number of
edges between the subsets is at most $\eps n$.} whose time complexity and probe complexity are $O(d^{\rho(\eps)})$, assuming that $\rho$ is known.
Minor-free graphs are a subclass of hyperfinite graphs (for an appropriate choice of $\rho$ that
depends on the size of the excluded minor).
In~\cite{LMRRS} the authors show that in every $f$-non-expanding graph\footnote{A graph is \emph{$f$-non-expanding} if every $t$-vertex subgraph $H$ satisfies $\phi_H \le f(t)$ where $\phi_G$ is the (edge) \emph{expansion} of $G$, that is, $\phi_G = \min_S \s{\partial_G(S)}/\s{S}$ where
the minimum is taken over all $S \subseteq V(G)$ of size $1 \leq |S| \leq |V(G)|/2$.}  $G$ where $f(t)=\Omega((\log t)^{-1}(\log\log t)^{-2})$ one can remove $\eps n$ edges from $G$
so that each connected component of the remaining graph is of size at most
$2^{2^{O(1/\eps)}}$. This enables them to provide a local sparse spanning graph algorithm for this family of graphs with probe complexity $d^{2^{2^{O(1/\eps)}}}$ and stretch $2^{2^{O(1/\eps)}}$.

Ram and Vicari~\cite{Rm2011} studied the problem of constructing sparse spanning graphs in the distributed model and provided an algorithm that runs in $\min\{D(G), O(\log n)\}$ number of rounds where $D(G)$ denotes the diameter of $G$.

The model of {\em local computation algorithms} as considered in
this work, was defined by Rubinfeld et al.~\cite{RTVX} (see
also Alon et al.~\cite{ARVX12}). Other local algorithms, for maximal
independent set, hypergraph coloring, $k$-CNF and maximum
matching include those given in~\cite{RTVX,ARVX12,MRVX12,MV13, EMR14}.

%% file: local-spanning-minor-free.tex
\section{Preliminaries}\label{sec:prel}

In this section we provide the precise definition of the algorithmic problem 
addressed in this paper, as well as several other useful definitions and notations.

We consider undirected, simple graphs over $n$ vertices, where the degree of each vertex is bounded
by $d$. We assume for simplicity that the set of vertices, $V$, is simply $[n] = \{1,\dots,n\}$,
so that there is a total order over the (identifiers of the) vertices.
For each vertex $v$, there is some arbitrary, but fixed order over its neighbours.
The input graph $G=(V,E)$ is given via an {\em oracle access to its incidence-list representation}.
Namely,
the algorithm is supplied with  $n$ and $d$, and has access to an oracle that for
any pair $(v,i)$ such that $v \in [n]$ and $i\in [d]$, the oracle either returns the
$i^{th}$ neighbour of $v$ (according to the aforementioned order over neighbours) or an indication that $v$ has less than $i$ neighbours. We refer to each such access to the oracle as a {\em probe\/} to the graph.
We now turn to formally define the algorithmic problem we consider in this paper.


\renewcommand{\labelenumi}{\theenumi}
\renewcommand{\theenumi}{\roman{enumi}.}

\BD 
\label{dfn:SSG-alg}
An algorithm $\calA$ is an {\em $(\e,q, \delta)$-local sparse spanning graph algorithm} if, given $n,d \ge 1$ and oracle access to the incidence-lists representation of a connected graph $G=(V,E)$ over $n$ vertices and degree at most $d$,
it provides
query access to a subgraph $G'=(V, E')$ of $G$ such that:
\BE
\item $G'$ is connected. 
\item $\s{E'} < (1+\eps)\cdot n$ with probability at least
    $1-\delta$ (over the
    internal randomness of $\mathcal{A}$).
\item\label{it:internal-rand} $E'$ is determined by $G$ and the internal randomness
    of $\mathcal{A}$.
\item $\calA$ makes at most $q$ probes to $G$.
\EE
By ``providing query access to $G'$'' we mean that
on input $\{u, v\}\in E$,
    $\calA$ returns whether $\{u, v)\} \in E'$ and for any
    sequence of queries, $\calA$ answers consistently with
    the same $G'$.

An algorithm $\calA$ is an {\em $(\e,q, \delta)$-local sparse spanning graph
algorithm for a family of graphs $\mathcal{C}$}
if the above conditions hold, provided that the input graph
$G$ belongs to $\mathcal{C}$.
\ED

\renewcommand{\labelenumi}{\theenumi}
\renewcommand{\theenumi}{\arabic{enumi}.}

We are interested in local algorithms that
for each edge they are queried on, perform
as few probes as possible to $G$. Ideally, we would like the number
of probes to be independent of $n$ and polynomial in $1/\eps$, $d$,
and possibly some parameters of the family $\mathcal{C}$.
We are also interested in bounding the total amount of space
used by the local algorithm, and its running time (in the word-RAM model).
Note that Item~\ref{it:internal-rand} implies that the answers of the algorithm to
queries cannot depend on previously asked queries.





We denote by $\d_G(u, v)$ (and sometimes by $\d(u, v)$ when $G$ is clear from the context) the distance between two vertices $u$ and $v$ in $G$.
We let $N(v)$ denote the set of neighbors of $v$ and for $\ell\geq 0$ let
$\Gamma_{\dnew{\ell}}(v) \eqdef \{ u\in V: \d(u, v) \leq \dnew{\ell})\}$.

Another parameter of interest is the stretch of $G'$. Given a connected graph $G=(V, E)$, a subgraph $G'=(V, E')$ is a $t$ -spanner of $G$ if for every $u,v \in V$, $\frac{\d_{G'}(u,v)}{\d_{G}(u,v)} \leq t$. In this case $t$ is referred to as the {\em stretch factor} of $G'$.

The total order over the vertices induces a total order (ranking) $r$ over
the edges of the graph in the following straightforward manner:
$r(\{u,v\}) < r(\{u',v'\})$ if and only if $\min\{u,v\} < \min\{u',v'\}$
or $\min\{u,v\} = \min\{u',v'\}$
and $\max\{u,v\} < \max\{u',v'\}$ (recall that $V = [n]$).
The total order over the vertices also induces an order over those vertices visited by
a Breadth First Search (BFS) starting from any given vertex $v$, and whenever we refer to
a BFS, we mean that it is performed according to this order.

Recall that a graph $H$ is called a {\em minor\/} of a graph $G$ if $H$ is isomorphic to a graph that can be obtained by zero or more edge contractions on a subgraph of $G$. A graph $G$ is {\em $H$-minor free} if $H$ is not a minor of $G$.
We will use the following theorem.
\BT[\cite{Mad68}]\label{thm:mader}
Let $c(s)$ be the minimum number $c$ such that every graph $G = (V, E)$ with $|E| \geq c\cdot|V|$ contracts to a complete graph $K_s$. Then $c(s) \leq 8s\log s$.
\ET

We shall use the following result from previous work (see Section~6 in ~\cite{ABI86}).  
\BT\label{thm:twise}
For every $1\leq t \leq n$, there exists an explicit construction of
a $t$-wise independent random variable $x = (x_1, \ldots, x_n)\in [q]^n$ for $q = \Theta(n)$
    whose seed length is at most $O(t \log n)$ bits.
Moreover, for any $1\leq i \leq n$, $x_i$ can be computed in  $O(t)$ operations in the word-RAM model.
\ET
\section{\dnew{The Algorithm}}\label{sec:upper}

In this section we prove the following theorem.

\BT\label{thm:main}
Algorithm~\ref{alg:ssg} is an $(\e,\poly(1/\eps,d,h),\delta)$-local sparse spanning graph algorithm
for graphs that are $H$-minor free, where $h$ is the size of $H$
and
$\delta = 1/\Omega(n)$. 
Furthermore, the stretch factor of $G'$ is $\tilde{O}(h \cdot \log d/\eps)$.
More precisely, the probe complexity of the algorithm is $\tilde{O}((h/\eps)^4 d^5)$.
Its space complexity (length of the random seed) is 
$\tilde{O}((h/\eps)d\log n)$ bits, and its running time is $\tilde{O}((h/\eps)^5 d^5)$ (in the Word-RAM model).
\ET


We begin by describing a global partition of the vertices. We  later describe how to locally generate this partition and 
\dnew{design our algorithm (and the sparse subgraph it defines), based on this partition.}

\subsection{The Partition $\calP$}\label{subsec:partition}
The partition described in this subsection, denoted by $\calP$, is a result of a random process.
We  describe how the partition is obtained in three steps where in each step we refine the partition from the previous step.
\dnew{The partition is defined based on three parameters: $\gamma \in (0,1)$,
an integer
$k>1$ and an integer $s > 1$, which is set subsequently (as polynomials of
$d$, $\eps$ and $h$).}

\ifnum\conf=1
\vspace{-1ex}
\fi
\paragraph{First Step.} We begin with some notation.
Given a subset of vertices $\dnew{\W} \subseteq V$
\dnew{and a vertex $v\in V$, we define the {\em center\/} of $v$ with respect to $\W$, denoted
$\cent(v)$ as the
vertex in $\W$ that is closest to $v$, breaking ties using vertex ids. That is,
$\cent(v)$ is the vertex with the minimum identifier
in the subset $\{y\in \W : \d(y, v) = \min_{w\in \W} \d(w,v)\}$.
}
For each $w\in \W$ we define the {\em cell\/} of $w$ with respect to $\W$ as
$\cell(w) \eqdef \{v\in V: \cent(v) = w \}$.
Namely, the set of vertices in $\cell(w)$ are the vertices which are closer to $w$ more than any other vertex in $W$ (where ties are broken according to the order of the vertices).
Notice that these cells form a partition of $V$.
Our initial partition is composed of these cells when picking $\W$ in the following way: each vertex $i\in [n]$ draws a $\gamma$-biased bit, denoted $x_i$, and $i\in W$ if an only if $x_i  = 1$. The joint-distribution of $(x_1, \ldots, x_n)$ is $t$-wise independent where $t \eqdef 2kd$.
(The reason that the choice of $\W$ is determined by a $t$-wise independent distribution rather than
an $n$-wise independent distribution is so as to bound the space used by the local emulation of
the global algorithm.)

\ifnum\conf=1
\vspace{-1ex}
\fi
\paragraph{Second Step.}
In this step we identify a subset of {\em special\/} vertices, which we call the {\em remote vertices\/} and make these vertices singletons \dnew{in the partition $\calP$}.
The set of remote vertices, $R$, is defined with respect to $\W$ and an integer parameter $k$ as described next.
Let $\dnew{\ell}_k(v)$ be the minimum integer $\ell$ such that the BFS tree rooted at $v$ of depth $\ell$ has size at least $k$.
Let $B_k(v)$ be the set of vertices in the BFS tree rooted at $v$ of depth $\ell_k(v)$.
We define $R = \{v\in V : B_k(v) \cap \W = \emptyset\}$, i.e., those vertices $v$ for which $B_k(v)$ does not contain a center.
Clearly, a vertex can identify efficiently if it is in $R$ by probing at most $kd$ vertices and checking whether they intersect $W$.
In Subsection~\ref{subsec:b-remote} we obtain a bound on the size of $R$.

\ifnum\conf=1
\vspace{-1ex}
\fi
\paragraph{Third Step.}
In this step we decompose cells that are still too big. We first argue that the cells are still connected \dnew{(after the removal the vertices in $R$ from all original cells)}. Thereafter we will use a procedure of tree decomposition in order to break the cells into smaller \dnew{parts}.
\BL\label{lem:shrt}
For every $w \in \W$, the subgraph induced by $\cell(w) \setminus R$ is connected.
Furthermore, for every $v \in \cell(w) \setminus R$, the subgraph induced \dnew{by} $\cell(w) \setminus R$ contains all vertices that belong to the shortest paths between $v$ and $w$.
\EL
\BPF
\dnew{Fix $w \in \W$, and consider any }
$v\in \cell(w) \setminus R$. We will prove that the subgraph induced by $\cell(w) \setminus R$ contains all \dnew{vertices on the  shortest paths} between $v$ and $w$ and this will imply the connectivity as well.
The proof is by induction on the distance to $w$.
In the base case $v =  w$. In this case $\cell(w) \setminus R$ clearly contains a path between $v$ to itself because it contains $v$.
Otherwise, we would like to show that for any $u\in N(v)$ for which $\d(u, w) < \d(v, w)$ it holds that $u \in \cell(w) \setminus R$.
The proof will then follow by induction.
let $\Path$ be a shortest path between $v$ and $w$ and let $\{v, u\}\in E$ denote the first edge in $\Path$.
We first observe that $\cent(u) = w$ and thus $u\in \cell(w)$.
Assume otherwise and conclude that
there is a vertex in $w' \in \W$ for which \dnew{either $\d(v,w') < \d(v,w)$ or}
$\d(v ,w) = \d(v, w')$ and $id(w') < id(w)$, in contradiction to the fact that $\cent(v) = w$ (see the definition of $\cent(\cdot)$ in the First Step).
Since $u$ is on a shortest path between $v$ and $w$ it follows that
\BEQ
\Gamma_{\d(u, w)-1}(u) \subseteq \Gamma_{\d(v, w)-1}(v)\;.\label{eq:gamma}
\EEQ
From the fact that $v \notin R$ it follows that $|\Gamma_{\d(v, w)-1}(v)| \leq k$ and hence from Equation~\eqref{eq:gamma} it follows that $|\Gamma_{\d(u, w)-1}(u)| \leq k$ and so $u \notin R$ as well.
We conclude that $u \in \cell(w) \setminus R$ and $\d(u, w) = \d(v , w) - 1$ as desired.
\EPF

We shall use the following notation. For each $w\in \W$ let $\calT(w)$ denote the BFS tree rooted at $w$ of the subgraph induced by $\cell(w) \setminus R$
\dnew{(recall that the BFS is performed by exploring the vertices according to the order of their identifiers (in $[n]$))}.
To obtain the final refinement of our partition, for each $w\in \W$  \dnew{such that} $|\calT(w)| > \dnew{\sz}$, we run Algorithm~\ref{alg:treed} on $\calT(w)$, $w$ and $\dnew{\sz}$.
\begin{algorithm}
{\small
\caption{{\bf (Recursive Tree decomposition)}}\label{alg:treed}
\textbf{Input:} A tree $\calT$, the root of the tree $v$ and an integer $\sz$.\\
\textbf{Output:} A decomposition of $\calT$ into subtrees, where each subtree is assigned a (sub-)center.
\BE
\item Initialize the set of vertices of the current part $Q :=\emptyset$.
\item \label{step:bfs} Perform a BFS starting from $v$ and stop \dnew{at} level $\ell \eqdef \ell_\sz(v)$ (see the definition of $\ell_\sz(\cdot)$ in the Second Step). Add to $Q$ all the vertices explored in the BFS.
\item Let $S$ denote the set of all the children of the vertices in the $\ell^{\rm th}$ level of the BFS (namely, all the vertices in level $\ell+1$).
\item For each vertex $u \in S$:
\BE
\item \label{step:check}If the subtree rooted at $u$, $\calT_u$, has size at least $\dnew{\sz}$, then disconnect this subtree from $\calT$ and continue to decompose by recursing on input $\calT_u$, $u$ and $s$.
\item Otherwise, add the vertices of $\calT_u$ to $Q$.
\EE
\item \label{step:q} Set $v$ to be the {\em sub-center} of all the vertices in $Q$.
\EE
}
\end{algorithm}

\subsection{The Edge Set}\label{subsec:edges}
Given the partition $\calP$ 
\dnew{defined in the previous subsection (Subsection~\ref{subsec:partition}),
we define the}
edge set of our sparse spanning graph $E'$ \dnew{in the following simple manner}.
In each part of $\calP$ which is not a singleton, we take a spanning tree. Specifically, we take the BFS-tree rooted at the sub-center of that part (see Algorithm~\ref{alg:treed}, Step~\ref{step:q}).
For every pair of parts of $X, Y \in \calP$, if the set
$\dnew{E(X,Y) \eqdef} \{\{x, y\}\in E : x\in X \text{ and } y \in Y\}$ is not empty, then we add to $E'$
\dnew{the edge $e \in E(X,Y)$ with minimal ranking (where the ranking over edges is as defined in
Section~\ref{sec:prel}).}
Clearly $G'$ is connected and spans $G$. We would like to bound the size of $E'$. To this end we will use \ifnum\conf=0
the following claim.
\else
the following simple claim.
\fi
\BCM\label{clm:part-size}
\dnew{The number of parts in $\calP$ is bounded as follows:}
$|\calP| \leq |\W| + |R| + \frac{n}{\sz}$.
\ECM
\ifnum\conf=0
\BPF
Consider the three steps of refinement of the partition.
Clearly, after the first step the size of the partition is exactly $|\W|$. After the second step, the size of the partition is exactly $|\W| + |R|$. Finally, since in the last step we break only parts
whose size is greater than $\sz$ into 
\dnew{smaller parts} that are of size at least $\sz$, we obtain that the number of new parts that are introduced in this step is at most $n/\sz$. The claim follows.
\EPF
\fi

The next lemma establishes the connection between the size of $\calP$ and the sparsity of $G'$.
\ifnum\conf=1
It readily follows from Theorem~\ref{thm:mader}. 
\fi
\BL\label{lem:Eprime-size}
For an input graph $G$ which is a $H$-minor free for a graph $H$ over $h$ vertices,
\ifnum\conf=0
$$|E'| 
   < n + |\calP| \cdot c(h)\;,$$
\else
 $|E'| 
   < n + |\calP| \cdot c(h)$,
\fi  
   where $c(h)$ is as defined in Theorem~\ref{thm:mader}.
\EL
\ifnum\conf=0
\BPF
Since for each $X\in \calP$ the subgraph induced by $X$ is connected, we can contract each part in $\calP$ and obtain an $H$-minor free graph.
The number of vertices in this graph is $|\calP|$. If we replace multi-edges with single edges, then by Theorem~\ref{thm:mader} we obtain that the number of edges in this graph is at most $ |\calP| \cdot c(h)$.
Finally, since the total number of edges in the union of spanning trees of each part it
$n-\dnew{|\calP| < n}$, we obtain the desired result.
\EPF
\fi

\subsection{Bounding the Number of Remote Vertices}\label{subsec:b-remote}

In this subsection we prove the following lemma.
\BL\label{lem:R-size}
If $k = \Omega( (\log^2(1/\gamma) + \log d)/\gamma)$, then
with probability at least $1-\frac{1}{\Omega(n)}$ it holds that $|R| \leq \gamma n$.
\EL

In order to establish Lemma~\ref{lem:R-size}
we start defining for every $v \in V$,
\begin{equation}
Y_k(v) \eqdef \{u\in V: v \in B_k(u) \}\;.
\label{eq:Ykv-def}
\end{equation}
Informally, $Y_k(v)$ is the set of vertices that encounter $v$ while performing a BFS which stops after the first level in which the total number of explored vertices is at least $k$.
We first establish the following simple claim.
\BCM\label{clm:Y-k}
For every vertex $u\in Y_k(v)$ and for every vertex $w$ that is on a shortest path
between $u$ and $v$, we have that $w \in Y_k(v)$.
\ECM
\BPF
Let $\ell = \d(u,v)$ and  $\ell' = \d(w,v)$, so that $d(u,w) = \ell-\ell' \geq 1$.
Assume, contrary to the claim, that $w \notin Y_k(v)$. This implies that
$|\Gamma_{\ell'-1}(w)| \geq k$. But since $\Gamma_{\ell'-1}(w) \subseteq \Gamma_{\ell-1}(u)$,
we get that $|\Gamma_{\ell-1}(u)| \geq k$, contrary to the premise of the claim that
$u \in Y_k(v)$.
\EPF

\medskip
We now turn to upper bound the size of $Y_k(v)$.
\BL \label{lem:struct}
For every graph $G = (V,E)$ with degree bounded by $d$, and for every $v\in V$,
\ifnum\conf=0
$$
|Y_k(v)| \leq d^3 \cdot k^{\log k+1}\;.
$$
\else
$|Y_k(v)| \leq d^3 \cdot k^{\log k+1}$.
\fi
\EL
\BPF
\dnew{Fix a vertex} $v\in V$.
For every $0 \leq j\leq k$, define $Y^j_k(v) \eqdef \{u\in Y_k(v):  \d(v, u) = j\}$.
Observe that $Y_k(v) = \bigcup_{j=0}^k Y^j_k(v)$.
Therefore, if we bound the size of $Y^j_k(v)$, for every $0 \leq j\leq k$, we will get a bound on the size of $Y_k(v)$.
\dnew{Consider first any} $3 \leq j < k$ and any vertex $u \in Y^j_k(v)$.
\dnew{Recall that $\ell_k(u)$ is the the minimum integer $\ell$ such that the BFS tree rooted at $v$ of depth $\ell$ has size at least $k$.}
Since $j \leq \ell_k(u)$, it follows that $|\Gamma_{j-1}(u)| < k$.
Now consider a shortest path between $u$ and $v$ and let $w$ be the vertex on this path for which
$\d(u, w) = \lfloor(j-1)/2\rfloor$.
Denote $q \eqdef \d(w, v)$.
By 
Claim ~\ref{clm:Y-k}, $w\in Y_k(v)$, and by the definition of $q$,
$w \in Y^q_k(v)$.
Therefore,
\BEQ
|\Gamma_{q-1}(w)| \leq k\;.\label{eq:ell}
\EEQ
From the fact that $w$ is on the shortest path between $u$ and $v$ it also follows that
\BEQN
q &=& \d(v,u) - \d(u, w)
= j -  \lfloor(j-1)/2\rfloor \nonumber \\
&=& \lceil(j-1)/2 \rceil + 1  \label{eq:ceil} \\
&\geq& \lfloor(j-1)/2\rfloor + 1 \nonumber
= \d(u, w) + 1\;.
\EEQN
Therefore $q- 1 \geq \d(u,w)$ and so $u \in \Gamma_{q-1}(w)$.
It follows that
\BEQ
Y^j_k(v) \subseteq \bigcup_{w\in Y^q_k(v)} \Gamma_{q-1}(w)\;.\label{eq:yjv}
\EEQ
From Equations~\eqref{eq:ell} and~\eqref{eq:yjv} we get that $|Y^j_k(v)| \leq k \cdot |Y^q_k(v)|$.
For every $j \leq 3$ we have the trivial bound that $|Y^j_k(v)| \leq d^3$.
By combining with Equation~\eqref{eq:ceil} we get that $|Y^j_k(v)| \leq d^3 \cdot k^{\log j}$. Since $Y_k(v) = \bigcup_{j=0}^k Y^j_k(v)$ we obtain the desired bound.
\EPF

\medskip
While the bound on $|Y_k(v)|$ in Lemma~\ref{lem:struct} may not be tight, it suffices for
our purposes. One might conjecture that it is possible to prove a polynomial bound (in $k$ and $d$).
We show that this is not the case
(see Lemma~\ref{lem:Y-k-big} in the appendix).

\smallskip
We now use the bound in Lemma~\ref{lem:struct} in order to bound the number of remote vertices.

\smallskip
\BPFOF{Lemma~\ref{lem:R-size}}
Let $\chiW$ denote the characteristic vector of the set $\W$.
For a subset $S \subseteq V$, let $\chiW(S)$ denote the projection of $\chiW$ onto $S$.
That is, $\chiW(S)$ is a vector of length $|S|$ indicating for each $x\in S$ whether
$\chiW(x) = 1$.

For each vertex $v\in V$ define a random variable $Z_v$ indicating whether it is a remote vertex with respect to $W$.
Recall that $v$ is remote if and only if $B_k(v) \cap \W = \emptyset$.
Recalling that $\W$ is selected according to a $t$-wise independent distribution
where $t = 2kd$ and that $k \leq |B_k(v)| < k\cdot d$, we get that $\Pr[Z_v =1] \leq (1-\gamma)^k$.
We also define $S_v \eqdef \{u\in V: B_k(u) \cap B_k(v) \neq \emptyset\}$.
Fix $v\in V$ and observe that the value of $Z_v$ is determined by $\chiW(B_k(v))$.
Furthermore, since for every $v\in V$ and $u \in V\setminus S_v$, $\chiW(B_k(v))$ and $\chiW(B_k(u))$ are independent it follows that $Z_u$ and $Z_v$ are independent as well.
Hence, in this case $\cov[Z_v, Z_u] = 0$, and we obtain the following upper bound
on the variance of the number of remote vertices.
\BEQN
\Var\left[\sum_{v \in V} Z_v\right] &=& \sum_{(v, u)\in V} \cov[Z_v, Z_u]
\;=\;  \sum_{v\in V} \sum_{u\in S_v} \left(\Exp[Z_v\cdot Z_u] - \Exp[Z_u]\cdot\Exp[Z_v]\right)\nonumber\\
&\leq& \sum_{v\in V} \sum_{u\in S_v} \Exp[Z_v \cdot Z_u |Z_v = 1]\cdot \Pr[Z_v = 1] 
\;\leq\; \sum_{v\in V} |S_v| \cdot (1-\gamma)^k\;.\label{eq:sv}
\EEQN
By the definition of  $Y_k(\cdot)$ in Equation~(\ref{eq:Ykv-def}) it follows that
$S_v \subseteq \bigcup_{u\in B_k(v)} Y_k(u)$.
By Lemma~\ref{lem:struct}, $\max_{v\in V}\{|Y_k(v)|\} \leq d^3\cdot k^{\log k +1}$.
Therefore
\BEQ
|S_v| \leq |B_k(v)| \cdot d^3\cdot k^{\log k +1} \leq d^4\cdot k^{\log k +2}\;.\label{eq:bkv}
\EEQ

Hence, by Equations~\eqref{eq:sv} and~\eqref{eq:bkv} we get
$\Var\left[\sum_{v \in V} Z_v\right] \leq n d^4\cdot k^{\log k +2} \cdot (1-\gamma)^k$. Since $(1-\gamma)^k \leq e^{-\gamma k}$ we obtain that $\Var\left[\sum_{v \in V} Z_v\right] \leq \gamma^2 n$  for $k = \Omega( (\log^2(1/\gamma) + \log d)/\gamma)$.
Since for every $v\in V$, $\Pr\left[Z_v = 1\right] \leq (1-\gamma)^k \leq \gamma$, we get that $\Exp\left[\sum_{v \in V} Z_v\right] \leq \gamma n/2$.
By applying Chebyshev's inequality we get that
\BEQ
\Pr\left[\sum_{v \in V} Z_v \geq \Exp\left[\sum_{v \in V} Z_v\right] + \gamma n/2\right]\leq \frac{4\Var\left[\sum_{v \in V} Z_v\right]}{\gamma^2 n^2} \leq \frac{4}{n}\;.\nonumber
\EEQ
Since $|R| = \sum_{v \in V} Z_v$ it follows that $|R| < \gamma n$ with probability at least $1-(4/n)$, as desired.
\EPFOF

\subsection{The Local Algorithm}
In this subsection we provide Algorithm~\ref{alg:ssg}, which on input $e\in E$, locally decides whether $e\in E'$, 
as defined in Subsection~\ref{subsec:edges}, based on the (random, but not completely
independent) choice of $\W$. 
 Given an edge $\{u,v\}$, the algorithm first finds, for each
$y \in \{u,v\}$, the center, $\cent(y)$, that $y$ is assigned to by the initial partition,
under the condition that $\cent(y) \in B_k(y)$. This is done by calling Algorithm~\ref{alg:findc},
which simply performs a BFS starting from $y$ until it encounters a vertex in $\W$, or it reaches
level $\ell_k(y)$ without encountering such a vertex (in which case $y$ is a remote vertex).
Algorithm~\ref{alg:findc} assumes access to $\chiW$, which is implemented using the random seed
that defines the $t$-wise independent distribution, and hence determines $\W$.
If $y$ is not found to be a remote vertex, then Algorithm~\ref{alg:ssg} next determines to which sub-part
of $\cell(\cent(y))\setminus R$ does $y$ belong. This is done by emulating the tree decomposition of the
BFS tree rooted at $\cent(y)$ and induced by $\cell(\cent(y))\setminus R$. A central procedure
that is employed in this process is Algorithm~\ref{alg:bfs}. This algorithm is given
a vertex $v\in \W$, and a vertex $u$ in the BFS subtree rooted at $v$ and induced
by $\cell(v)\setminus R$. It returns the edges going from $u$ to its children in this tree,
by performing calls to Algorithm~\ref{alg:findc}.

\begin{algorithm}
{\small
\caption{{\bf (Sparse Spanning Graph)}}\label{alg:ssg}
\textbf{Input:} An edge $\{u,v\} \in E$. \\
\textbf{Output:} YES if $\{u,v\} \in E'$ and NO otherwise.
\BE
\item For each $y\in \{u, v\}$ find the part that $y$ belongs to as follows:
\BE
\item Use Algorithm~\ref{alg:findc} to obtain $\cent(y)$.
\item If $\cent(y)$ is `null' then the part that $y$ belongs to is the singleton set $\{y\}$.
\item Let $\calT$ denote the BFS tree rooted at $\cent(y)$ in the subgraph induced by $\cell(\cent(y)) \setminus R$. By Lemma~\ref{lem:shrt} every shortest path between $\cent(y)$ and $v\in \cell(\cent(y)) \setminus R$ is contained in the subgraph induced on $\cell(\cent(y)) \setminus R$. Therefore the edges of $\calT$ can be explored via Algorithm~\ref{alg:bfs}.
\item Reveal the part (the subset of vertices) that $y$ belongs to in $\calP$ (as defined in Subsection~\ref{subsec:partition}). Recall that the part of $y$ is the subtree of $\calT$ that contains $y$ after running Algorithm~\ref{alg:treed} on input $\calT$, $\cent(y)$ and $\sz$. This part can be revealed locally as follows.  \label{step:part}
\BE
\item Reveal the path between $\cent(y)$ and $y$ in $\calT$, denoted by $\Path(y)$. Since $\Path(y)$ is the shortest path between $y$ and $\cent(y)$ with the lexicographically smallest order it can revealed by performing a BFS from $y$ until $\cent(y)$ is encountered.
\item Run Algorithm~\ref{alg:treed} on $\calT$ while recursing in Step~\ref{step:check} only on the subtrees in which the root is contained in $\Path(y)$.
\EE
\EE
\item If $u$ and $v$ are in the same part,
then return YES iff the edge $\{u, v\}$ belongs to the BFS tree of that part.
\item Otherwise, return YES iff the edge $\{u, v\}$ is the 
\dnew{edge with minimum rank connecting the two parts}.
\EE
}
\end{algorithm}

\medskip
\BPFOF{Theorem~\ref{thm:main}}
We set $\gamma = \eps/2c(h)$, $k = \Theta( (\log^2(1/\gamma) + \log d)/\gamma)$, and
$\sz = c(h)/4\eps$.
We start by bounding the size of $\W$ (with high probability).
By the definition of $\W$ we have that
$\Exp[|\W|] = \Exp\left[ \sum_{i\in [n]} \chiW(i)\right] = \gamma n$.
Since for every $1 \leq i <  j \leq n$, $\chiW(i)$ and $\chiW(j)$ are pairwise-independent,
\ifnum\conf=0
we obtain that
$$\Var\left[\sum_{i\in [n]} \chiW(i)\right] = \sum_{i\in [n]} \Var \left[\chiW(i)\right] = \sum_{i\in [n]} \left(\Exp \left[\chi^2_\W(i)\right] - \Exp \left[\chiW(i)\right]^2\right) = n\gamma(1-\gamma)\;.$$
Therefore,
\fi
by Chebyshev's inequality
$\Pr \left[|W| \geq 2\gamma n\right] \leq \frac{1-\gamma}{\gamma n}$.
By Lemma~\ref{lem:R-size} (and the setting of $k$), with probability $1-1/\Omega(n)$,
$|R| \leq \gamma n$.
By Claim~\ref{clm:part-size}, Lemma~\ref{lem:Eprime-size} and the settings of
 $\gamma$ and $\sz$, we get that $|E'| \leq (1+\eps)n$ with probability $1-1/\Omega(n)$.

The claim about the stretch of $G'$ follows from the fact that the diameter of every part of $\calP$ is bounded by $2k$.
\ifnum\conf=1
We defer the details of the bound on the probe complexity and running time to Appendix~\ref{app:probe-ssg}.
\else

The number of vertices that Algorithm~\ref{alg:findc} inspects
(for any vertex $v$ it is called with) is at most $kd$. Since the degree of each vertex is bounded by $d$, its probe complexity is bounded by $kd^2$.
Algorithm~\ref{alg:findc} makes at most $kd$ accesses to $\chiW$, hence, by Theorem~\ref{thm:twise} its running time is bounded by $O(k^2d^2)$.
The probe complexity of Algorithm~\ref{alg:bfs} is upper bounded by the total probe complexity
of at most $d$ calls to  Algorithm~\ref{alg:findc}, plus $d$ executions of a BFS until at most
$kd$ vertices are encountered (Steps~\ref{st:u-up} and~\ref{st:up-w}, which for simplicity
we account for separately from Algorithm~\ref{alg:findc}).
A similar bound holds for the running time.
Hence, the total probe complexity and running time  of Algorithm~\ref{alg:bfs} are  $O(k d^3)$ and $O(k^2 d^3)$, respectively.

The size of any subtree returned by Algorithm~\ref{alg:treed} is upper bounded by $\sz^2  d^2$.
To verify this, recall that at the end of Step~\ref{step:bfs} of Algorithm~\ref{alg:treed}, at most $\sz d$ vertices were explored. Hence, the number of vertices that are incident to the explored vertices is at most  $\sz d^2$.
Thus, due to Step~\ref{step:check}, the total number of vertices in each part is at most $\sz^2 d^2$.
Since Step~\ref{step:check} can be implemented locally by calling Algorithm~\ref{alg:bfs} at most $\sz$ times, we obtain that the total probe complexity and running time of revealing each part is at most $O(\sz^2 k d^5)$ and $O(\sz^2 k^2 d^5)$, respectively.
The probe complexity and running time of Algorithm~\ref{alg:ssg} are dominated by those of Step~\ref{step:part}.
Observe that in Step~\ref{step:part} at most $|\Path(y)|$ parts are revealed.
Since $|\Path(y)|  \leq k$, we obtain that  the overall probe complexity and running time  of Algorithm~\ref{alg:ssg} are bounded by $O(\sz^2 k^2 d^5)$ and $O(\sz^2 k^3 d^5)$, respectively.
By the settings of $k$ and $\sz$ we obtain the final result.
\fi
\EPFOF

\begin{algorithm}
{\small
\caption{{\bf (Find Center)}}\label{alg:findc}
\textbf{Input:} A vertex $v$ and an integer $k$. Query access to $\chiW$.\\
\textbf{Output:} $\cent(v)$ if $\cent(v) \in B_k(v)$ or `null' otherwise.
\BE
\item Perform a BFS from $v$ until
\dnew{the first level that contains a vertex in $\W$ or
until at least $k$ vertices are reached. That is, defining $\W_j \eqdef \Gamma_j(v) \cap \W$, stop at}
level $\ell$ where
$\ell \eqdef \min\{ \ell_k(v), \min_{j}\{{\W_j \neq \emptyset}\}  \}$.
\item If  $\W_\ell = \emptyset$ then return  \textbf{`null'} ($v$ is remote).
\item Otherwise, return the vertex with the minimum id from $\W_\ell$.
\EE
}
\end{algorithm}

\begin{algorithm}
{\small
\caption{{\bf (get BFS outgoing-edges endpoints)}}\label{alg:bfs}
\textbf{Input:} $v\in \W$ and a vertex $u \in \dnew{\calT(v)}$ where $\calT(v)$ denotes the BFS tree induced by $\cell(v) \setminus R$ and rooted at $v$.\\
\textbf{Output:} The outgoing edges from $u$ in $\calT(v)$ (the orientation of the edges is from the root to the leaves).
\BE
\item Initialize $S: = \dnew{\{v\}}$  
\item For each $u' \in N(u)$, if 
 \dnew{the following three conditions} hold, then add $u'$ to $S$:
\BE
\item Algorithm~\ref{alg:findc} on input $u'$ returns $v$. Namely, $\cent(u') = v$.
\item \label{st:u-up} $u$ is on a shortest path between $u'$ and $v$. Namely, $\d(u', v)= \d(u, v) + 1$.
\item \label{st:up-w}
$u$ is the vetex with the minimum id among all vertices in  $\{w \in N(u'): \d(w, v) = \d(u', v) - 1\}$.
\EE
\item Return $S$.
\EE
}
\end{algorithm}